\documentclass[preprint,12pt,sort&compress]{elsarticlemod}
\usepackage[margin=1.5in]{geometry}
\usepackage{rotating}
\usepackage{lineno}
\usepackage{overpic}
\usepackage{color}
\usepackage[multiple]{footmisc}
\usepackage{caption,subcaption}
\usepackage{tikz}
\usepackage{natbib}
\usepackage{comment}
\usepackage{hyperref}
\usepackage{amsmath}

\journal{Nucl. Instrum. and Methods Phys. Res. A, 161683}
\begin{document}
\begin{frontmatter}

\title{End-user experience with the SCoTSS Compton imager \\and directional
  survey spectrometer}

\author[NRcan,CUP]{Laurel E. Sinclair}\ead{laurel.sinclair@canada.ca}
\author[NRcan]{Andrew McCann}\ead{andrew.mccann@canada.ca}
\author[NRC,CUP]{{Patrick R.B. Saull}}
\author[NRcan,NRC,CUP]{Nathan Murtha}
\author[NRC]{Rodger L. Mantifel}
\author[NRC]{Christian V.O. Ouellet}
\author[DRDC]{Pierre-Luc Drouin }
\author[NRC]{Audrey M.L. Macleod}
\author[CBSA]{Brian Le Gros}
\author[RCMP]{Ian Summerell}
\author[RSI]{Jens H. Hovgaard}
\author[RSI]{Stephen Monkhouse} 
\author[RSI]{Flaviu Stanescu}
\author[DRDC]{Guy Jonkmans}

\address[NRcan]{Canadian Hazards Information Service, Natural Resources Canada}
\address[CUP]{Department of Physics, Carleton University}
\address[NRC]{Measurement Science and Standards, National Research Council Canada}
\address[CBSA]{Canada Border Services Agency}
\address[RCMP]{Royal Canadian Mounted Police}
\address[RSI]{Radiation Solutions Inc.}
\address[DRDC]{Defense Research and Development Canada}

\begin{abstract}
The Silicon photomultiplier-based Compton Telescope for Safety and Security
(SCoTSS) has been developed incorporating end-user requirements into the
design process. The end-user group includes those responsible for mobile
survey in the event of a radiological or nuclear accident, those responsible
for radiation survey support to security operations at major events and at
Canadian borders, as well as some of those responsible for Canadian defense
applications.  The SCoTSS development program has reached a
technology readiness level of eight, and we are proceeding with field trials
of the instrument in high-fidelity operational environments. Prospective end
users have been involved in trial set up and execution, assuring applicability
in their mission spaces. 
SCoTSS has been subject to trials involving hidden
sources, heavily shielded sources, imager moving with respect to source, 
and complicated man-made surroundings. 
Our operators value high sensitivity for anomaly geolocation and 
mapping.
End users also require an instrument which is capable of direction 
reconstruction in motion, as well as rapid
imaging of a field of view. 
We have developed a ``time to image'' measure which allows for quantitative
comparison of imagers of fundamentally different technology, where one design
may have an advantage in terms of energy resolution and compactness and 
another design may have an advantage in terms of efficiency and cost effectiveness.
We present here the performance of the SCoTSS imager in rapid direction
finding.
As well, we compare the time to image quantity for the SCoTSS imager and the 
H3D Polaris-H Quad imager where the data were taken under equivalent
conditions.
This quantitative measure of imaging performance can allow
operators to make an informed choice of the design that meets their needs
taking into consideration also weight and size as well as budgetary constraints.
\end{abstract}


\end{frontmatter}








\section{Introduction}
The Silicon photomultiplier-based Compton Telescope for Safety and Security, SCoTSS, is the first
Compton imager based on solid scintillators coupled to silicon photomultiplier
(SiPM) optical sensors
\cite{2014ITNS...61.2745S}. SCoTSS is composed of pixelized layers of crystal
scintillator in a modular configuration allowing users to specify the mass
required for their operations.  SCoTSS has been designed for use in radiological and nuclear
incident investigations, nuclear non-proliferation applications, and to support
safety and security activities in Canada. A full nine-module imager in a 
3~x~3-module configuration employs a total scintillator mass
of over 15~kg.
Thus, the $3\times3$-module version of SCoTSS can act as a survey spectrometer with the sensitivity to do isotope
identification and alarming at dose rates consistent with background levels in
sub-second acquisitions.
Such sensitivity enables 
mobile survey operations in total-count mode with real-time alarming,
isotope identification, and geo-location.
In addition,
SCoTSS can act as a directional survey spectrometer, with source localization
anywhere within the entire 4$\pi$ solid angle based on 
real-time self-shielding indicators.
Once a radiation field of interest has alarmed, the operator can switch from isotope alarming mode to targeting specific
fields and directions for gamma-ray imaging.
A smaller single-module version of the SCoTSS imager can be utilized where compactness
is a priority for hand-held or backpack use.

In this work we review the performance of the SCoTSS $3\times3$-module imager
with specific reference to the requirements of operational deployment.
Several other gamma imagers are now available commercially, some also based on
Compton imaging~\cite{PHDs,H3D}, and some on coded-aperture or pinhole mask 
techniques~\cite{Canberra}.
Security and emergency response operators require quantification of imager
performance in order to choose the optimum device for their operations.
We define a quantity called ``time-to-image'' which can be used to quantitatively compare the
performance of any imager designs, provided data are acquired under equivalent
conditions.  Data under equivalent conditions have been acquired for the
SCoTSS $3\times3$-module imager and single-module imagers and for the commercially available H3D Polaris-H
Quad Compton
imager~\cite{2015NIMPA.784..377W}.
These three devices will be compared herein.
Other groups can reproduce these experimental conditions in the future and
calculate time-to-image as well to bring more imagers into the comparison.

\section{The SCoTSS Instrument}
\subsection{Hardware Design}
The parameters of the final design of the SCoTSS imager were chosen after several years of
prototyping and simulation activity
\cite{2009ITNS...56.1262S,2010SPIE.7665E..1ES}.
Crystal scintillator was chosen as the active medium in
order to be able to provide the sensitivity and field-worthiness of standard
gamma-ray mobile survey and mapping instruments currently in use by radiological-nuclear
security teams.
In specific, thallium-doped cesium iodide was chosen for its ease of use and for 
the correspondence of its 
optical emission spectrum to the spectral response of the 
SensL~\cite{SensL} silicon
photomultipliers.
 The SCoTSS imager consists of
two segmented layers. 
An array of cubic crystals,
each 1.35~cm on a side define
the front face of the detector. The parameters of this front ``scatter'' layer
were optimized to ensure Compton scatter interactions yield excellent
positional and energy resolution, while minimizing the probability of multiple
scatters within the layer. The second ``absorber'' layer, an array of cubic
CsI(Tl) crystals 2.8~cm on a side, has been optimized to absorb all of the remaining
energy of gamma rays scattering in the front layer while achieving the
required level of positional resolution. Each SCoTSS crystal is individually
mated to an SiPM sensor and optically isolated.

The SCoTSS instrument employs
a modular design, with a single module consisting of 
a $4\times4$ array of crystals
in both the scatter and absorber layers.  This design allows for
different configurations of SCoTSS modules to be combined and arranged
depending on what is suitable and practical in a given deployment setting.

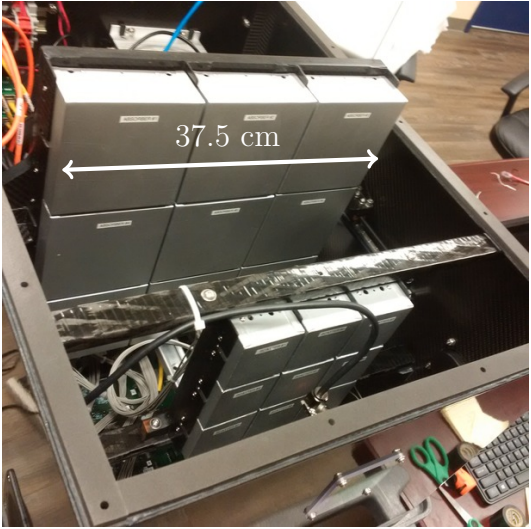
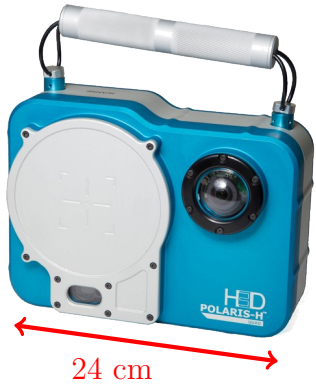
\begin{figure}[htb]
\centering
\begin{subfigure}[t]{0.5\textwidth}
\centering
\begin{tikzpicture}
    \node[anchor=south west,inner sep=0] at (0,0) {\includegraphics[width=\textwidth]{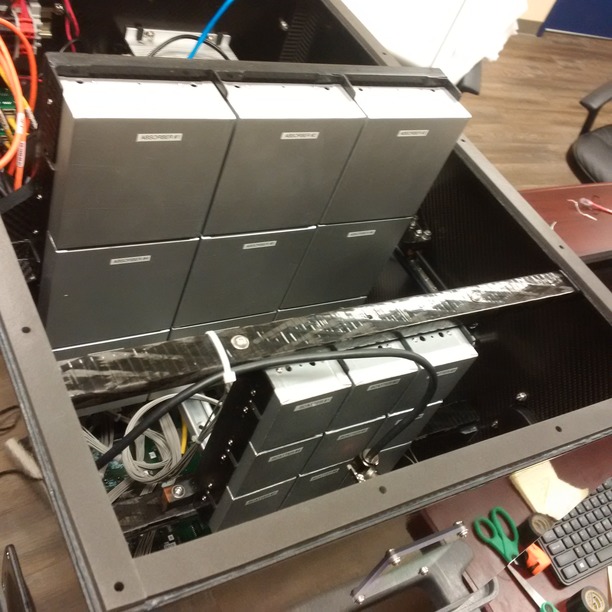}};
    \draw[white,ultra thick,<->] (.8,4.75) -- (5.,4.9);
    \node[white,ultra thick] at (3,5.2) {37.5 cm};
\end{tikzpicture}\caption{SCoTSS Imager}
\end{subfigure}%
\begin{subfigure}[t]{0.5\textwidth}
\centering
\begin{tikzpicture}
    \node[anchor=south west,inner sep=0] at (0,0) {\includegraphics[width=\textwidth]{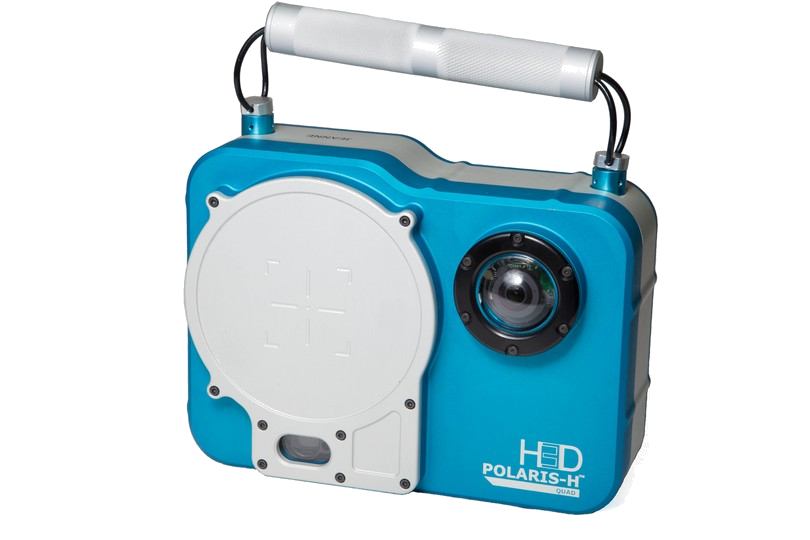}};
    \draw[red,ultra thick,<->] (1.7,0.38) -- (5.2,-0.1);//for two-column
    \node[red,ultra thick] at (3,-0.2) {24 cm};//for two-column
\end{tikzpicture}\caption{H3D Polaris-H Quad Imager}
\end{subfigure}
\caption{Photographs of the imagers.  (a) $3\times3$-module SCoTSS Imager with the top panel of the
  enclosure removed. The $3\times3$ arrays of single SCoTSS modules are
  visible. The smaller array is the scatter layer and the larger array is the
  absorber layer.  A dimension of 37.5~cm across the absorber layer is
  indicated.  This is an irreducible dimension for this design whereas the
  exterior housing is still under development.  (b) H3D Polaris-H Quad
  commercially available imager with indication of longest exterior dimension.}
\label{fig:photos}
\end{figure}

Figure~\ref{fig:photos}~(a) shows the $3\times3$-module SCoTSS imager discussed here.
It employs a $3\times3$ configuration of single
modules, resulting in $12\times12$ arrays of
measurement channels in both the scatter and absorber layers.

Readout of the channels is done by custom-designed coincidence-timing and
digitizing electronics connected to the SiPM sensors. An optical camera,
positioned behind a transparent panel in the front of the imager enclosure,
enables overlay of gamma-ray and optical images. Inside the carbon-fibre
enclosure, the scatter and
absorber layers are attached to rails which allow the distance
separating the layers to be adjusted.  This ``nuclear zoom'' feature 
allows for an initial low resolution wide-angle field of view image to be
followed by a higher precision image over a narrower field of 
view.
For this study the inter-plane distance in both simulation and data was set 
to 20~cm and 8~cm for the $3\times3$-module and single-module imagers respectively.

The SCoTSS imager in a $3\times3$ configuration is suitable for truckborne
mobile survey applications.  A single module can also be assembled for hand-held 
or backpack deployments. The Polaris-H Quad imager is a commercially available
imager from H3D which is also suitable for hand-held application.  It is pictured in 
Figure~\ref{fig:photos}~(b).  The Polaris-H Quad imager utilizes cadmium zinc
telluride semiconductor technology and weighs 3.5~kg.

\subsection{Sensitivity and Energy Resolution}
\begin{figure}[htb]
\centering
\includegraphics[width=0.85\textwidth]{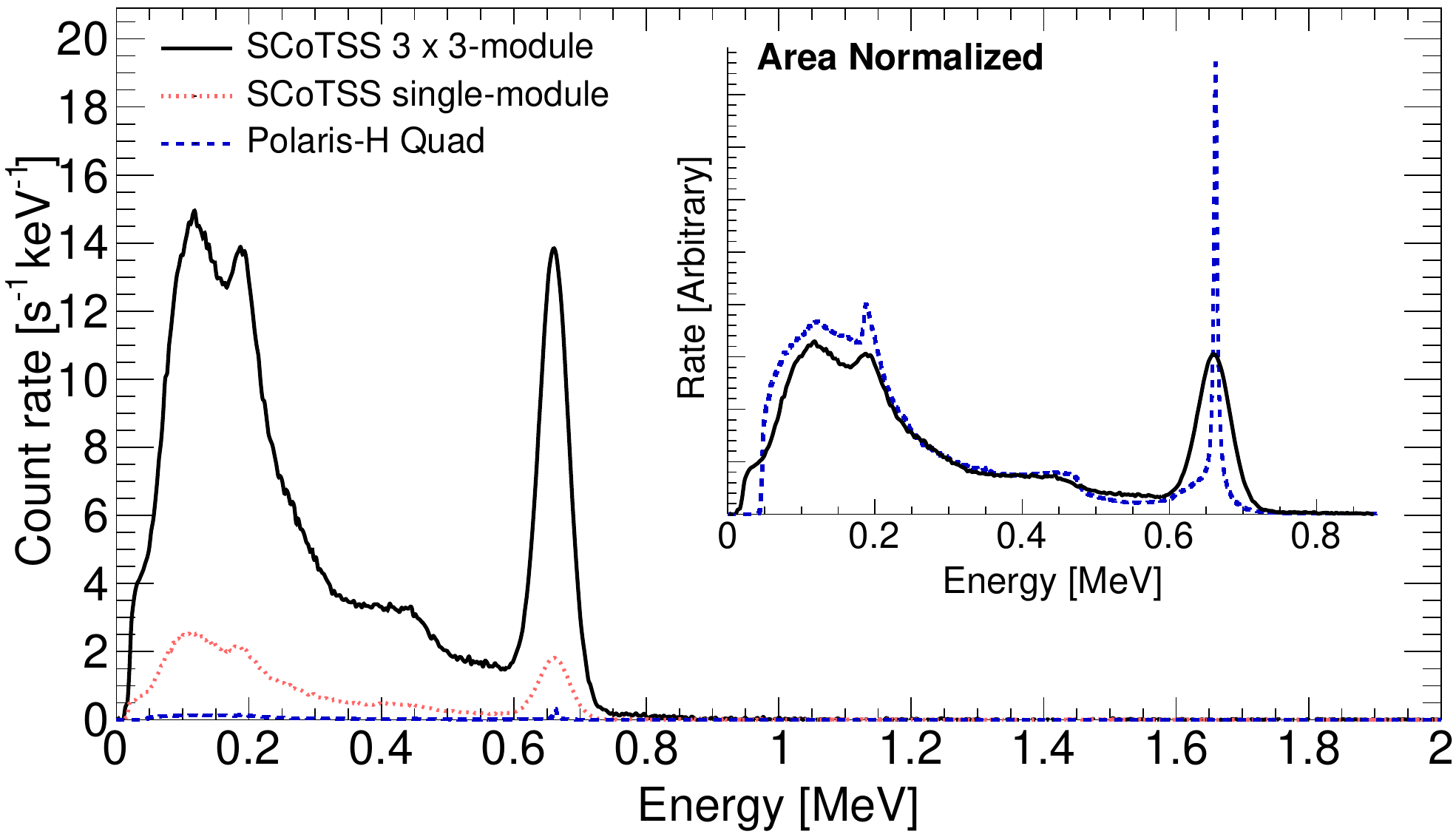}
\caption{Energy spectra measured by the SCoTSS and the H3D Polaris-H Quad
  imagers when exposed to a gamma-ray flux equivalent to that emitted by a
  1~mCi Cs-137 source located at a distance of 10~m for one second. 
(Multiple one second spectra were averaged to produce these smooth
  distributions.) The
  insert, zoomed over the $0-0.88$~MeV energy range, shows the spectra
  area-normalized and serves to illustrate the relative energy resolution of
  the two instruments. Some small high-energy features visible in the SCoTSS spectrum are
  due to imprecise background subtraction. The Polaris-H
  Quad spectrum was accumulated using a 5$\times$ stronger field and scaled
  down for the comparison, so natural background contributions to the 
  Polaris-H
  Quad spectrum
  are considered to be negligible.}
\label{fig:energy_spec}
\end{figure}

Figure~\ref{fig:energy_spec} shows the
average energy spectrum in counts per second measured by the
$3\times3$-module and single-module SCoTSS imagers when exposed to a 1~mCi Cs-137
source located at a distance of 10~m.
Evident in Figure~\ref{fig:energy_spec} is the 662~keV gamma-ray
emission line from Cs-137, illustrating the $\sim$7\% full-width at half
maximum (FWHM) energy resolution
of the SCoTSS imager at that energy.
The single-module SCoTSS imager accumulates a few counts per second in the
photopeak for this source which gives a dose rate at the detector of only
15~nGy/hr, less than the natural background.
The $3\times3$-module SCoTSS detector on the other hand accumulates around a hundred counts in
the photopeak in one second.
A one-second sampling rate is the typical rate used when performing high
sensitivity mobile radiometric survey and mapping~\cite{mobile_survey}.
Thus, the high sensitivity of the $3\times3$-module SCoTSS imager allows for isotope
identification and alarming of sources at background levels during truckborne survey.

The Cs-137 energy spectrum measured by the H3D
Polaris-H Quad instrument under equivalent conditions is also plotted in
Figure~\ref{fig:energy_spec} allowing a clear comparison of the sensitivity.
The Polaris-H Quad detector accumulates less than one count in the photopeak
per second under these conditions, thus it would require a much longer time of acquisition 
than either SCoTSS detector in
order to detect this source at this distance.

The inset figure in Figure~\ref{fig:energy_spec} shows the energy spectra of
the SCoTSS $3\times3$-module and Polaris-H Quad imagers area normalized.  This
allows for a comparison of the shapes of the spectra and it is clear that the 
cadmium zinc telluride detector used in Polaris-H Quad provides a superior energy
resolution.  Polaris-H Quad is quoted to achieve $\leq$1.1\% FWHM at 
662~keV~\cite{H3D_www_Jul2018}. 
There may be circumstances such as environments cluttered with many
radioactive sources of different energies of emission in which superior energy
resolution would provide improved imaging capability.  Exploration of these
effects will be the subject of future work. 

The self-shielding of the crystal configuration in the SCoTSS imagers can be utilized to calculate a
direction to a source from any incoming angle, without requring a Compton coincidence and without restriction to
a field of view.
The sensitivity of the $3\times3$-module imager means that these directions can be
calculated for relatively weak sources on a second by second basis.
Thus, from an operational standpoint, the $3\times3$-module SCoTSS can be deployed in a moving
vehicle (truck, helicopter, etc) which can survey a region and alarm when in
the field of a source of interest with a real time directional indicator that localizes
the direction of incoming radioactivity in 4$\pi$.

\subsection{Real-time display}
\begin{sidewaysfigure}[!htb]
\begin{subfigure}[t]{0.63\textwidth}
\begin{tikzpicture}
  \node[anchor=south west,inner sep=0] at (0,0) {\includegraphics[width=0.89\textwidth]{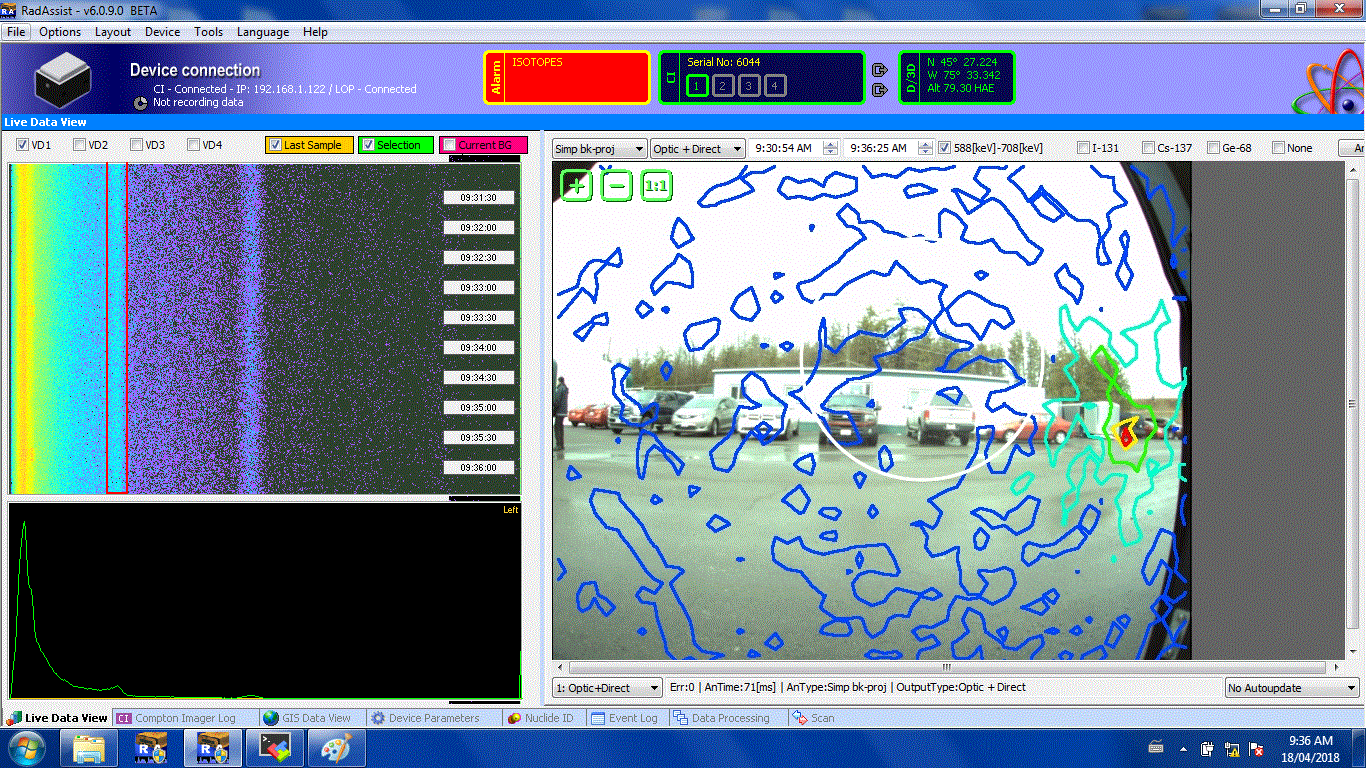}};
  \draw[white, ultra thick,<-] (1.,2.5) -- (1.3,2.0);
  \node[anchor=north west,inner sep=0, white, text width=3.2cm] at (1.0,1.9)
       {Time and energy range selection from one-second spectra.};
\end{tikzpicture}
\vspace{-.3cm}\caption{Screen grab of the real-time SCoTSS acquisition software.}
\end{subfigure}%
\begin{subfigure}[t]{.3\textwidth}
\centering
\hspace*{-.4cm}
\begin{tikzpicture}
  \node[anchor=south west,inner sep=0] at (0,0)
       {\includegraphics[width=1.2\textwidth]{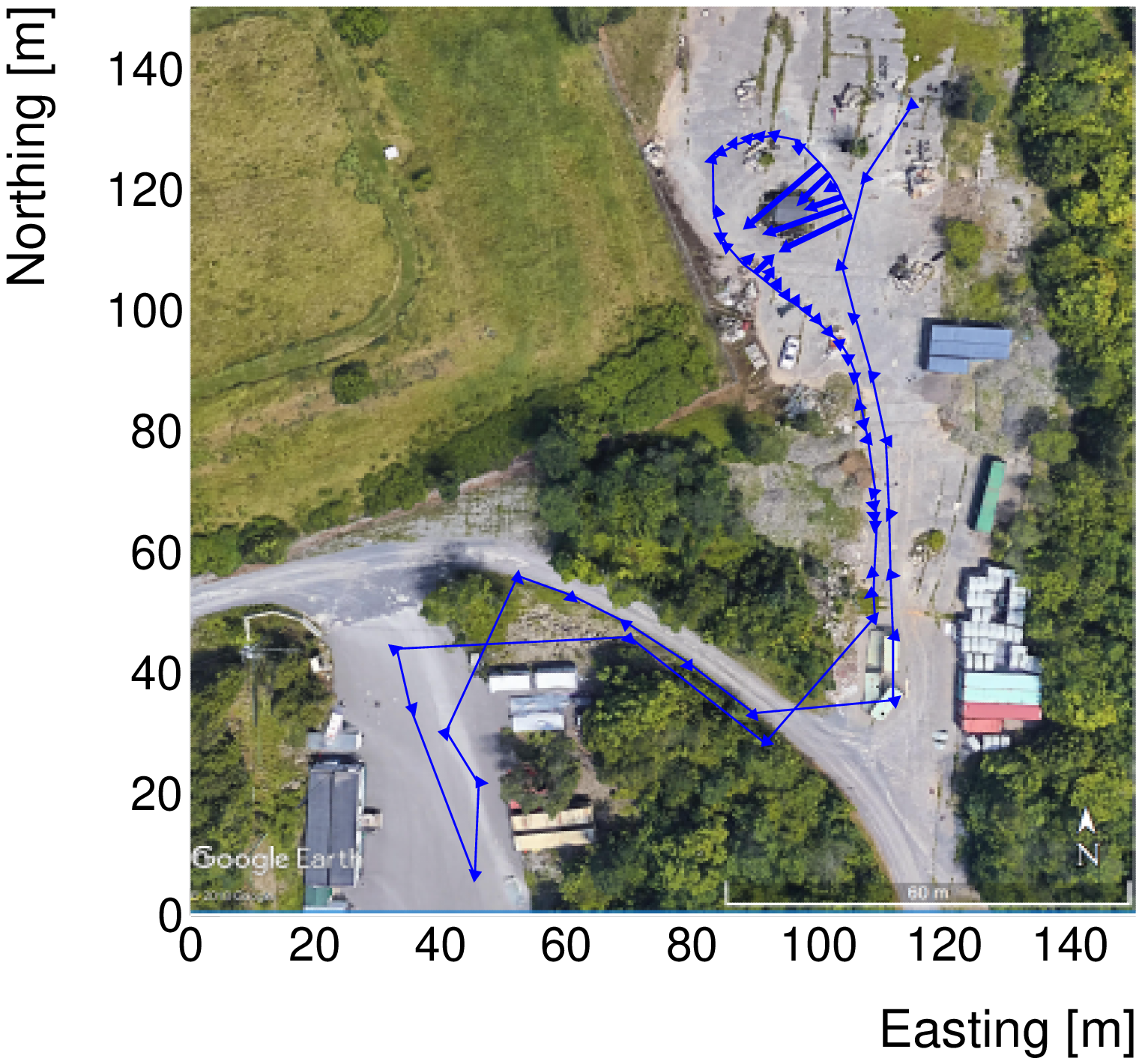}};
\end{tikzpicture}\vspace{-.3cm}
\caption{Geolocated bearing directions.}
\end{subfigure}
\caption{Real time information available to SCoTSS operator.
(a) The
  upper-left panel shows the one-second spectra ``waterfall'' plot, with the
  summed spectrum histogram plotted in the lower-left panel. The domain of the
  spectra is from 0~MeV to 3.072~MeV.  The panels are
  clickable with energy and time ranges selectable from the waterfall
  plot. The optical image in the right-hand panel is overlayed with contours
  from a Compton-cone back-projection algorithm showing the
  gamma-ray image corresponding to the time and energy ranges selected.
  The white
  circle in the image indicates the most likely direction of a source based on
  a five-second reconstruction using the self-shielding Simple Direction
  Finding (SDF) algorithm
  discussed in the text.
(b)  Blue arrows show the source bearing direction vectors obtained during a mobile survey with the SCoTSS imager and survey
spectrometer.
Each arrow was calculated using five seconds of data and applying the SDF
algorithm.  The arrows
are plotted with tail at the location of the imager and arrow length is
proportional to the count rate at that location.}
\label{fig:screen_grab}
\end{sidewaysfigure}

Figure~\ref{fig:screen_grab}~(a) shows an example of the screen available to the SCoTSS operator in real time.
The upper left windows shows a ``waterfall'' plot.  One-second energy spectra
are displayed with energy on the x-axis and time on the y-axis.  The number of
counts in each spectrum is indicated by the colour scale.
The lower left subwindow shows a total deposited energy spectrum, summed over
all detector channels.
By default the current energy spectrum accumulated in one second intervals is shown.
Alternately (as shown here), the user may select a region of time of interest on the waterfall plot, and
then the lower left subwindow will show the energy spectrum for that time
period.

The right subwindow shows the image from the optical camera with the contours
of the gamma image overlain.  The operator has selected an energy band
of interest from the waterfall plot and it is this energy band for which the
Compton image has been reconstructed.  Thus, it is not necessary for the
operator to know a priori the identity of the radioactive material of
interest in order to image it.
A white circle (difficult to see against the sky in this image) also indicates the result
of an independent assessment of source location based on simple self-shielding
direction finding using five seconds of data.

This particular image was grabbed from the SCoTSS operator laptop screen during an
operational exercise.  The source which is being imaged had not yet been
deployed.  It was situated within its heavily shielded shipping container in a
trailor well to the right side just within the imager's field of view.  The source was about
30~m away, on the opposite side of the parking lot from the location of the imager.

Figure~\ref{fig:screen_grab}~(b) shows a sequence of directions
reconstructed using the self-shielding information during an operational
exercise.  
The directions have
 been plotted in absolute coordinates.  The tail of each vector is located at
 the position of the survey vehicle containing the imager and the vectors
 indicate the compass direction to the source from that location.
From the standpoint of an operator, this rapid compass direction which allows
for optimal positioning of the survey vehicle is highly 
valuable in addition to the information contained within the Compton images.

\section{Methods}
\subsection{Self-shielding and directionality}

Though designed as a Compton imaging telescope, the SCoTSS imager can be
considered as a close-packed array of self-shielding elements. Self-shielding
directional methods use the relative rates of energy deposits in each detector element
to determine the most likely direction to the source of emission. 
Direction-finding results for a single-module imager making use of a look-up table method have been 
presented previously~\cite{McCann et al.(2018)}.
Here we will present direction-finding results for the $3\times3$ module
imager using a method which we call Simple Direction Finding (SDF).

\begin{figure}[htb]
\centering
\begin{subfigure}[t]{0.445\textwidth}
\centering
\begin{tikzpicture}
  \node[anchor=south west,inner sep=0] at (0.5,0.2) {\includegraphics[trim={0.5cm 0.43cm 13.9cm 0.445cm},clip,width=0.86\textwidth]{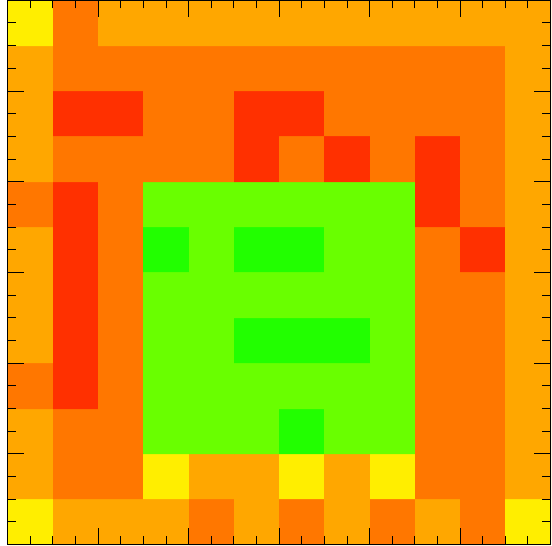}};
  \draw[ultra thick,<-] (0.6,0.0) -- (1.0,0.0);
  \node[ultra thick] at (1.3,0.0) {+x};
  \draw[ultra thick,->] (0.23,3.6) -- (0.23,4.0);
  \node[ultra thick] at (0.23,3.3) {+y};
\end{tikzpicture}\vspace{-0.2cm}\caption{Polar 170$^\circ$, Azimuthal 90$^\circ$}
\end{subfigure}%
\begin{subfigure}[t]{0.445\textwidth}
  \centering
\begin{tikzpicture} 
  \node[anchor=south west,inner sep=0] at (0.5,0.2) {\includegraphics[trim={0.5cm 0.43cm 13.9cm 0.445cm},clip,width=0.86\textwidth]{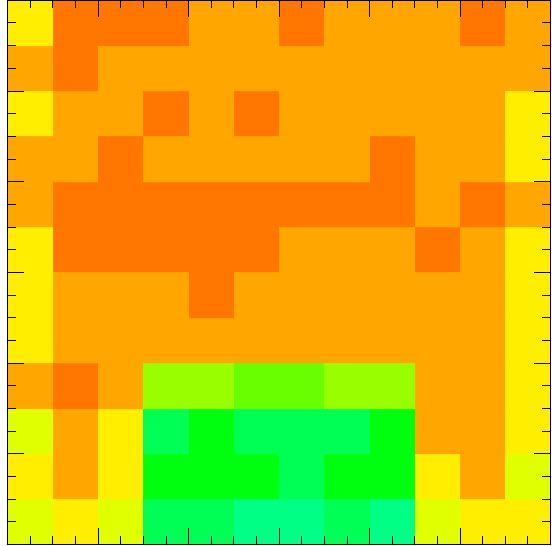}};
  \draw[ultra thick,<-] (0.6,0.0) -- (1.0,0.0);
  \node[ultra thick] at (1.3,0.0) {+x};
  \draw[ultra thick,->] (0.23,3.6) -- (0.23,4.0);
  \node[ultra thick] at (0.23,3.3) {+y};
\end{tikzpicture}\vspace{-0.2cm}\caption{Polar 140$^\circ$, Azimuthal 90$^\circ$}
\end{subfigure}
\begin{subfigure}[t]{0.445\textwidth}
\vspace{0.4cm}
\centering
\begin{tikzpicture}
  \node[anchor=south west,inner sep=0] at (0.5,0.2) {\includegraphics[trim={0.5cm 0.43cm 13.9cm 0.445cm},clip,width=0.86\textwidth]{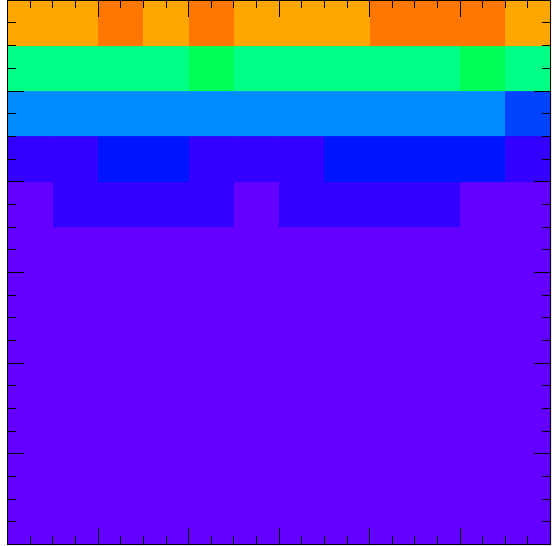}};
  \draw[ultra thick,<-] (0.6,0.0) -- (1.0,0.0);
  \node[ultra thick] at (1.3,0.0) {+x};
  \draw[ultra thick,->] (0.23,3.6) -- (0.23,4.0);
  \node[ultra thick] at (0.23,3.3) {+y};
\end{tikzpicture}\vspace{-0.2cm}\caption{Polar 90$^\circ$, Azimuthal 90$^\circ$}
\end{subfigure}%
\begin{subfigure}[t]{0.445\textwidth}
\vspace{0.4cm}
\centering
\begin{tikzpicture}
  \node[anchor=south west,inner sep=0] at (0.5,0.2) {\includegraphics[trim={0.5cm 0.43cm 13.9cm 0.445cm},clip,width=0.86\textwidth]{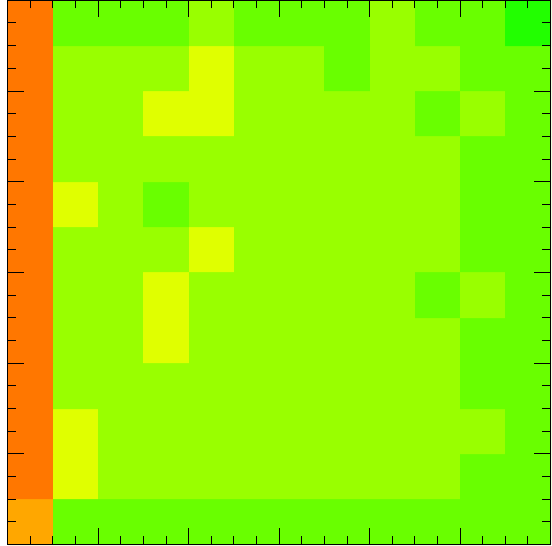}};
  \draw[ultra thick,<-] (0.6,0.0) -- (1.0,0.0);
  \node[ultra thick] at (1.3,0.0) {+x};
  \draw[ultra thick,->] (0.23,3.6) -- (0.23,4.0);
  \node[ultra thick] at (0.23,3.3) {+y};
\end{tikzpicture}\vspace{-0.2cm}\caption{Polar 60$^\circ$, Azimuthal 0$^\circ$}
\end{subfigure}
\begin{subfigure}[t]{0.7\textwidth}\centering
\vspace{0.3cm}
 \includegraphics[angle=270,width=\textwidth]{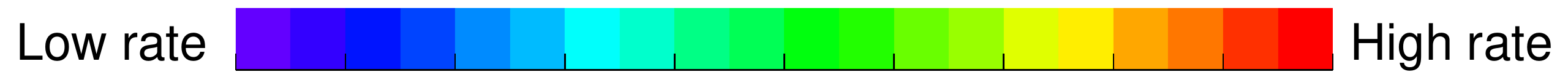}
\end{subfigure}
\caption{Self-shielding properties of the SCoTSS imager. Plotted is the
  relative count rates in the 12~$\times$~12 channel array of the SCoTSS
  absorber plane when a Cs-137 point source is located at the polar and
  azimuthal positions listed in the sub-captions. The low-rate regions in
  panels (a) and (b) indicate the shielding of the absorber plane by the
  material in the scatter plane. In panels (c) and (d), the source is located
  at the side of the detector, thus the material on the outer edges of the
  absorber shield the remaining material. Similar shielding effects occur in
  the scatter plane. Rates have been obtained from SCoTSS Geant4 simulations.
  Positive $z$-axis points into the page. See
  Figure~\ref{fig:imager_diagram} for a definition of angles and directions.}
\label{fig:shielding_rates}
\end{figure}

\begin{figure}[htb]
\centering \includegraphics[trim={1.9cm 4.1cm 16.6cm 5cm},clip,width=0.7\textwidth]{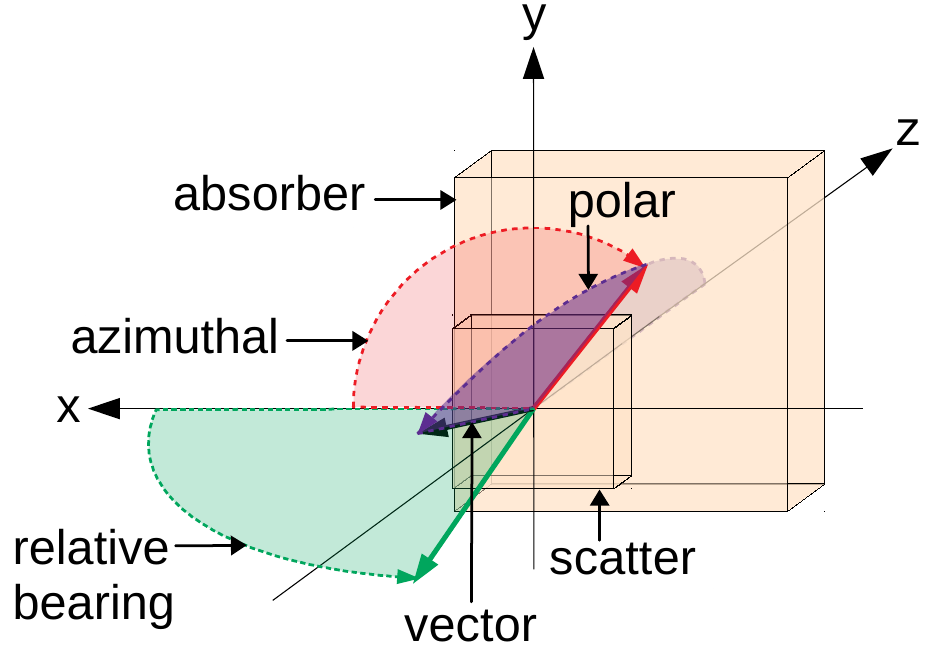}
\caption{Schematic diagram of the SCoTSS imager and coordinate
  system. Directional vectors relative to an origin at the centre of the front
  face of the scatter layer are defined using spherical coordinates with a
  polar angle (drawn in purple) between the $z$-axis and the vector
  and an azimuthal angle (drawn in red) between the $x$-axis and the
  vectors projection on the $xy$-plane. It is useful to define the
  relative bearing angle between the $x$-axis and the vector's
  projection on the $xz$-plane (drawn in green).} 
\label{fig:imager_diagram}
\end{figure}

Some of the self-shielding properties of
the $3\times3$-module SCoTSS imager are presented in Figure~\ref{fig:shielding_rates} where
the coordinate system and angle definitions are shown in
Figure~\ref{fig:imager_diagram}.
Figure~\ref{fig:shielding_rates} shows the relative count-rates of each crystal
in the 12~$\times$~12 channel array of the absorber layer for various positions
of a Cs-137 point source as determined through 
Geant4~\cite{2003NIMPA.506..250A} simulations.
In Figure~\ref{fig:shielding_rates}~(a) the source is at the front of the imager 
slightly above the horizon.  The scatter array thus casts a shadow on
central channels of the absorber layer.
In Figure~\ref{fig:shielding_rates}~(b) the source is 40$^\circ$ above the
horizon, and the shadow of the scatter plane on the absorber plane is that
much lower.
Figure~\ref{fig:shielding_rates}~(c) shows the source edge-on to the absorber
layer and the strong shielding of the entire layer results in straightforward 
reconstruction of the source direction in this case.
In Figure~\ref{fig:shielding_rates}~(d) the source is again toward the side but the
source may be reconstructed to be offset
somewhat from the edge since count rates are consistent with gamma rays also incident on the face of the absorber layer.

The SDF method starts
by choosing three perpendicular axes ($\hat{u},\hat{v},\hat{w}$) in the
detector geometry, e.g. ``top-bottom'', ``left-right'' and ``back-front''. The
difference in the measured rate projected along these directions is then
computed, resulting in a three-dimensional vector. If constructed sensibly,
this vector provides a crude estimate of the direction of the source of
emission. The SDF algorithm as applied to the SCoTSS $3\times3$-module imager is described in
Figure~\ref{fig:SDF}.
\begin{figure}[htb]
\centering
\begin{tikzpicture}[scale = 0.8]
\draw [ultra thin, green] (1*0.166,0) -- (1*0.166,2);
\draw [ultra thin, green] (2*0.166,0) -- (2*0.166,2);
\draw [ultra thin, green] (3*0.166,0) -- (3*0.166,2);
\draw [ultra thin, green] (4*0.166,0) -- (4*0.166,2);
\draw [ultra thin, green] (5*0.166,0) -- (5*0.166,2);
\draw [ultra thin, green] (6*0.166,0) -- (6*0.166,2);
\draw [ultra thin, green] (7*0.166,0) -- (7*0.166,2);
\draw [ultra thin, green] (8*0.166,0) -- (8*0.166,2);
\draw [ultra thin, green] (9*0.166,0) -- (9*0.166,2);
\draw [ultra thin, green] (10*0.166,0) -- (10*0.166,2);
\draw [ultra thin, green] (11*0.166,0) -- (11*0.166,2);
\draw [ultra thin, green] (0,1*0.166) -- (2,1*0.166);
\draw [ultra thin, green] (0,2*0.166) -- (2,2*0.166);
\draw [ultra thin, green] (0,3*0.166) -- (2,3*0.166);
\draw [ultra thin, green] (0,4*0.166) -- (2,4*0.166);
\draw [ultra thin, green] (0,5*0.166) -- (2,5*0.166);
\draw [ultra thin, green] (0,6*0.166) -- (2,6*0.166);
\draw [ultra thin, green] (0,7*0.166) -- (2,7*0.166);
\draw [ultra thin, green] (0,8*0.166) -- (2,8*0.166);
\draw [ultra thin, green] (0,9*0.166) -- (2,9*0.166);
\draw [ultra thin, green] (0,10*0.166) -- (2,10*0.166);
\draw [ultra thin, green] (0,11*0.166) -- (2,11*0.166);

\draw [ultra thin, green] (1*0.166 + 3,0) -- (1*0.166 + 3,2);
\draw [ultra thin, green] (2*0.166 + 3,0) -- (2*0.166 + 3,2);
\draw [ultra thin, green] (3*0.166 + 3,0) -- (3*0.166 + 3,2);
\draw [ultra thin, green] (4*0.166 + 3,0) -- (4*0.166 + 3,2);
\draw [ultra thin, green] (5*0.166 + 3,0) -- (5*0.166 + 3,2);
\draw [ultra thin, green] (6*0.166 + 3,0) -- (6*0.166 + 3,2);
\draw [ultra thin, green] (7*0.166 + 3,0) -- (7*0.166 + 3,2);
\draw [ultra thin, green] (8*0.166 + 3,0) -- (8*0.166 + 3,2);
\draw [ultra thin, green] (9*0.166 + 3,0) -- (9*0.166 + 3,2);
\draw [ultra thin, green] (10*0.166 + 3,0) -- (10*0.166 + 3,2);
\draw [ultra thin, green] (11*0.166 + 3,0) -- (11*0.166 + 3,2);
\draw [ultra thin, green] (0+3,1*0.166) -- (2+3,1*0.166);
\draw [ultra thin, green] (0+3,2*0.166) -- (2+3,2*0.166);
\draw [ultra thin, green] (0+3,3*0.166) -- (2+3,3*0.166);
\draw [ultra thin, green] (0+3,4*0.166) -- (2+3,4*0.166);
\draw [ultra thin, green] (0+3,5*0.166) -- (2+3,5*0.166);
\draw [ultra thin, green] (0+3,6*0.166) -- (2+3,6*0.166);
\draw [ultra thin, green] (0+3,7*0.166) -- (2+3,7*0.166);
\draw [ultra thin, green] (0+3,8*0.166) -- (2+3,8*0.166);
\draw [ultra thin, green] (0+3,9*0.166) -- (2+3,9*0.166);
\draw [ultra thin, green] (0+3,10*0.166) -- (2+3,10*0.166);
\draw [ultra thin, green] (0+3,11*0.166) -- (2+3,11*0.166);

\draw (0,0) --(0,2) -- (2,2) -- (2,0) -- (0,0);
\draw (1,0) --(1,2);
\draw (0,1) --(2,1);

\draw (3,0) --(3,2) -- (5,2) -- (5,0) -- (3,0);
\draw (4,0) --(4,2);
\draw (3,1) --(5,1);

\draw [thick, <->] (7,1-.5) -- (7,0-.5) -- (6,0-.5);
\node at (6.65,0.95-.5) {$\hat{y}$};
\node at (6,0.3-.5) {$\hat{x}$};
\node at (7.2,-.5) {$\otimes ~\hat{z}$};

\draw [thick, <->] (5.793,2.107) -- (6.5,1.4) -- (7.207,2.107);
\node at (5.793,2.307) {$\hat{u}$};
\node at (7.207,2.307) {$\hat{v}$};
\node at (6.8,1.45) {$\otimes ~\hat{w}$};

\node at (1,2.25) {Scatter};
\node at (0.5,0.5) {$S_2$};
\node at (1.5,1.5) {$S_1$};
\node at (0.5,1.5) {$S_0$};
\node at (1.5,0.5) {$S_3$};

\node at (4,2.25) {Absorber};
\node at (3.5,0.5) {$A_2$};
\node at (4.5,1.5) {$A_1$};
\node at (3.5,1.5) {$A_0$};
\node at (4.5,0.5) {$A_3$};
\end{tikzpicture}
\vspace{-0.5cm}
\begin{align}
 u_S =\,& \alpha ( S_0 - S_3 )   & u_A =\,& A_0 - A_3 &  \\
 v_S =\,& \alpha ( S_1 - S_2 )   & v_A =\,& A_1 - A_2 & \vspace{-3mm}  \\[-1.1em]
 u   =\,& u_A + u_S                 & v  =\,& v_A + v_S  & w =& \sum_{i=0}^{3}(A_{i} - \alpha S_{i})\\
\theta =\,&\arccos \frac{w}{\sqrt{u^{2} + v^{2} + w^{2}}} \span \span &\phi =& \arctan(v/u) - \pi/4
\end{align}
\vspace{-0.3cm}
\caption{Simple direction finding algorithm applied to the SCoTSS
  $3\times3$-module imager. The
  12~$\times$~12 channels in the scatter plane are grouped into four quadrants and
  summed, resulting in a 2~$\times$~2 signal array $S_{i=0,1,2,3}$. The same
  grouping is done in the absorber plane, resulting in the signal array
  $A_{i=0,1,2,3}$. The three-dimensional coordinate system
  $(\hat{u},\hat{v},\hat{w})$ is defined and the difference in the measured
  rates projected along these directions is computed, yielding the vector
  $(u,v,w)$ using the equations on lines 1, 2 and 3. A scaling parameter,
  $\alpha$, is introduced to weight the rates in the scatter plane and account
  for the relative difference in size and stopping power of the scatter plane compared to
  the absorber. The polar angle, $\theta$, and azimuthal angle, $\phi$, as
  defined in Figure~\ref{fig:imager_diagram} are computed using the standard
  Cartesian-to-spherical-polar equations on line 4. }
\label{fig:SDF}
\end{figure}
The geometry of the detector is first simplified by
grouping the rates in the scatter plane into 4 quadrants. The same is done for
the absorber plane, reducing the number of rates from 288 to eight. 
The $\hat{u}$
and $\hat{v}$ directions were chosen to lie in the $xy$-plane as defined in
Figure~\ref{fig:imager_diagram}, but with a $45^{\circ}$ rotation about the
$z$-axis. The $\hat{w}$ direction is parallel to the ``back-front'' direction
of the SCoTSS geometry which also lies parallel to the $z$-axis as defined in
Figure~\ref{fig:imager_diagram} ($\hat{w}\equiv \hat{z}$). Due to the difference in
size and mass between the scatter and absorber planes a scale factor weighting
the scatter rates, $\alpha$, is introduced when projecting the rate-difference
along the $\hat{w}$ direction. For the current study we set
$\alpha \equiv 6.8$. Using the standard Cartesian-to-spherical-polar
conversions (equation line 4 in Figure~\ref{fig:SDF}) the SDF vector is
converted into the polar and azimuthal angles defined in
Figure~\ref{fig:imager_diagram}.
 
\subsection{Compton Imaging}
Compton imaging methods require coincident energy depositions, reconstructing the
directional information of only those impinging gamma rays which scatter in the first layer
and are fully absorbed in the second. Though this requirement reduces the
number of considered detections to $\sim$0.5\% of the total number incident for
mid-range energies of $\sim$662~keV, those
which meet the selection criteria contain a high quality of directional
information. The incident direction of each detection is known to lie on the
surface of a cone projected outward from the detector along the axis
connecting the scatter and absorption interaction points. Successive cones
``back-projected'' outward from the detector intersect and, when overlayed,
form an image of the locations of emission. In the case of a single point source, only
three Compton coincidence events are needed to uniquely define the location of the source within
measurement uncertainty. Assuming a point source, a $\chi^2$-minimization
algorithm can be applied to the back-projected Compton cones to find the
direction most consistent with all cones~\cite{2009ITNS...56.1262S}. This method
allows for rigorous treatment of the measurement uncertainty on the individual
cones in the reconstruction, permitting a source position measurement with
high precision and accuracy.

To quantify the point-source localization performance of an instrument
or algorithm, we propose a metric called ``time-to-image'', which
is the reconstruction precision obtained after a certain acquisition
time \cite{2012NIMPA.679...89S}. The reconstruction precision is
formally the root mean square (RMS) spread of reconstructed Compton image
directions computed from independent datasets of a certain acquisition
time. Generally, as the acquisition time increases, greater and
greater angular precision is achieved. The basic instrument
characteristics: Doppler broadening, efficiency, sensitivity, energy and angular
resolution, combine and affect the time-to-image metric, which
can be seen as a global performance metric. Detector design
optimized for any one or any set of the basic instrument
characteristics, at the expense of others, can degrade the
time-to-image performance. Since the time-to-image metric can be
constructed for any instrument or algorithm which can localise a
point source it can be used to compare the performance of detectors which may
each have superiority in one or more of the individual performance measures.

\section{Results}
\subsection{Self-shielding and directionality}

The SCoTSS detector has a natural coordinate system as shown in
Figure~\ref{fig:imager_diagram}, with a pole along the symmetry axis connecting
the centres of the scatter and absorber planes.
In the natural coordinate system there is a polar angle with respect
to this pole, and an azimuthal angle about it.
As demonstrated in
Figure~\ref{fig:shielding_rates}, there is plenty of information in the pattern
of energy deposits from the inclusive (non-coincident) trigger, for an accurate reconstruction
of the azimuthal angle of a source direction about the symmetry axis.
That being said, the $3\times3$-module SCoTSS detector configuration was not optimized for self-shielding
directionality indicators. For example, Geant4 simulations (not shown) have 
shown that
when the scatter and absorber planes
are separated by 20~cm, the detector responses due to sources
located at polar angles near $60^\circ$ and $120^\circ$ are practically indistinguishable. In
these locations, neither plane shields the other from the incoming radiation
and the self-shielding pattern within the planes cannot break the
degeneracy in any reasonable length of time.
Note however that these simulations of the $3\times3$-module SCoTSS detector neglected much
of the passive detector hardware which may create shielding effects sufficient
to break or appreciably mitigate these degeneracies.  This will be
exploited in future work.

In any case, neither the polar nor the azimuthal angle with respect to the natural SCoTSS
coordinate system are of primary importance in
operations.
The angle of primary use in operations is the bearing angle, or the compass
direction to the source location from the detector's current position.
The bearing angle guides the vehicle operator closer to the source and
enables the operator to bring the source within the imager's field of view.
Relative bearing comes from the projection of the direction vector into the
$xz$-plane, the plane of the earth neglecting topography and curvature.
The relative bearing angle depends on both the azimuthal and
polar angles, as shown in
Figure~\ref{fig:imager_diagram}.
Absolute bearing angle is then calculated by adding the heading of the vehicle carrying
the imager to the relative bearing angle.

\begin{figure}[htb]
\centering
\includegraphics[width=0.7\textwidth]{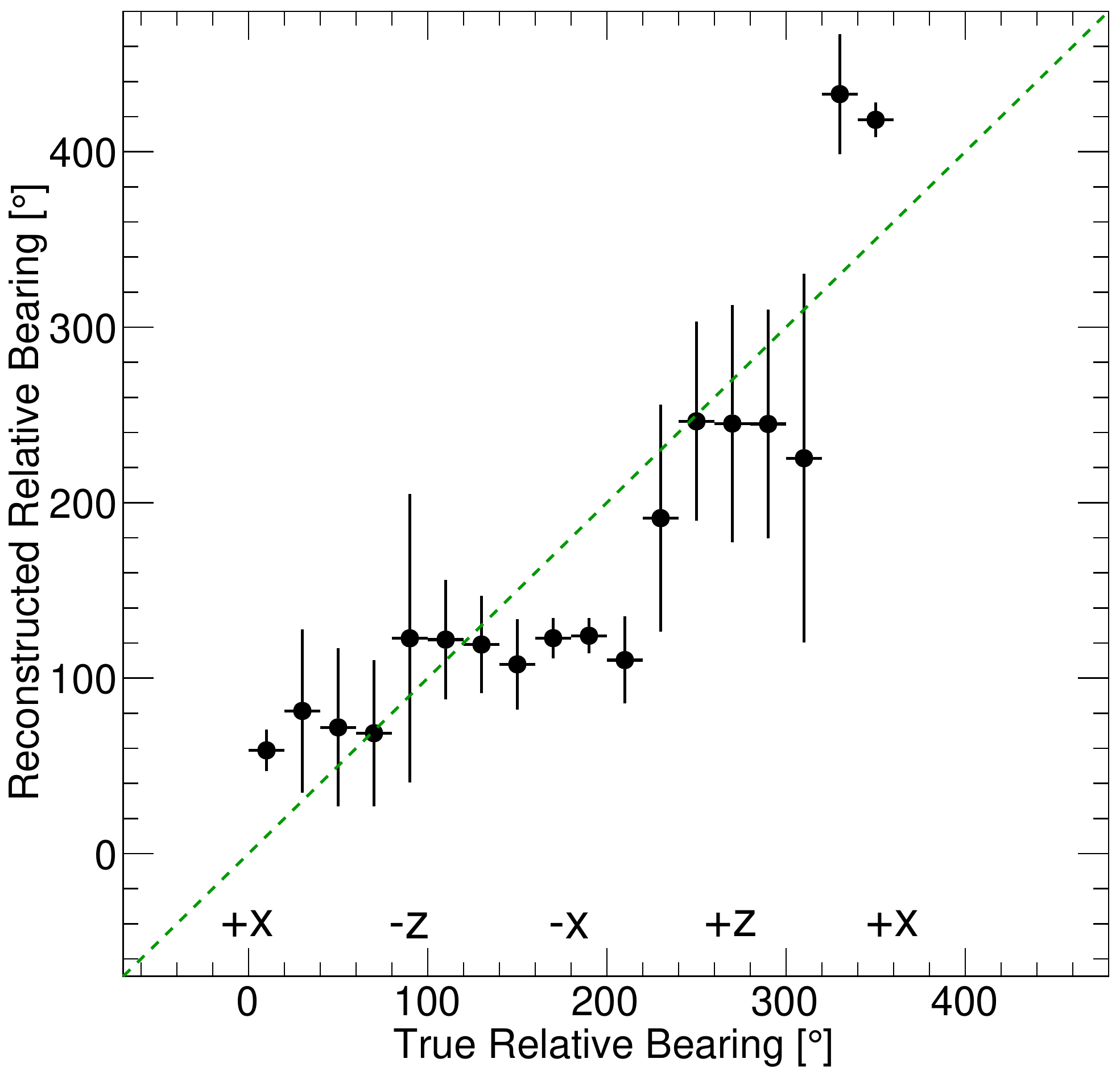}
\caption{Reconstructed relative bearing from the self-shielding SDF algorithm
  applied to Geant4 simulated data from the $3\times3$-module imager plotted against true relative
  bearing. 
The response of the $3\times3$-module SCoTSS imager to
  a 10~mCi Cs-137 point source located at a distance of 10~m 
 for trials of one second of data is shown. From each data set a relative
  bearing in the $xz$-plane is reconstructed, and the ensemble of
  reconstructed and true pairs is plotted using a profile histogram showing the
  mean and spread in each bin (black circles). The green dashed line indicates
  a perfect reconstruction ($y=x$).  See
  Figure~\ref{fig:imager_diagram} for a definition of angles and directions.}
\label{fig:bearing}
\end{figure}

The directional
response of the $3\times3$-module SCoTSS imager to
  a 10~mCi Cs-137 point source located at a distance of 10~m 
was calculated using the Geant4 simulation package.
Multiple trials were simulated at various polar and azimuthal directions
each with one second of data. From each data set a relative
  bearing in the $xz$-plane was reconstructed.
Figure~\ref{fig:bearing} shows the SDF reconstructed relative bearing against true relative
bearing for these one second trials.
The ensemble of
  reconstructed and true pairs is plotted using a profile histogram showing the
  mean and spread in each bin (black circles).
We find that, even for just a one-second acquisition period, there is a rough correspondence between the reconstructed and
true relative bearings.
Poor correlation arises in some regions largely due to degeneracy of the
self-shielding detector response for sources at some polar angles as mentioned
previously.

Self-shielding information can be calculated from every energy deposit,
without having to wait for a good Compton coincidence event.  Thus, the SDF
direction to the source can be
mapped while the instrument is in motion, even for relatively weak sources.
The utility of this angle in a live-agent field exercise was illustrated by
Figure~\ref{fig:screen_grab}~(b).
Despite the regions of poor correlation shown in Figure~\ref{fig:bearing} 
around relative bearing
of 200$^\circ$, the bearing angle information is of great use operationally.
It provides a first clue for
operators as to which side of the vehicle a threat agent is 
located on. 
This allows for orientation of the imager with the source within the field of
view, and rapid Compton imaging.

\subsection{Compton imaging}
\begin{figure}[htb]
\centering
\includegraphics[width=0.9\textwidth]{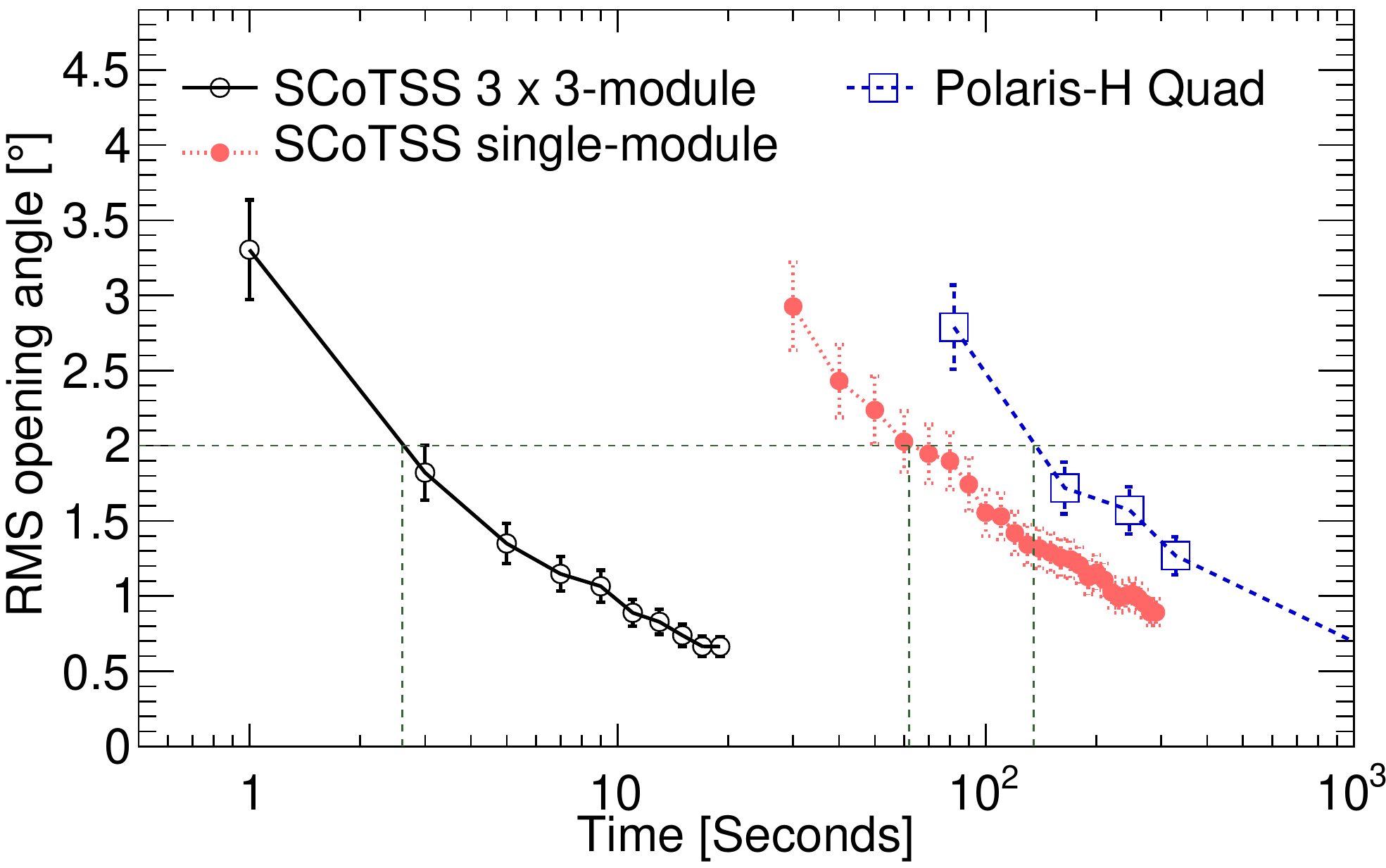}
\caption{Source reconstruction precision as a function of acquisition time for
  the SCoTSS and the H3D Polaris-H Quad imagers when exposed to a gamma-ray
  flux equivalent to that emitted by a 1~mCi Cs-137 point source located at a
  distance of 10~m and offset by $10^\circ$ from the negative $z$-axis (polar
  angle $170^\circ$).  A localization precision within $2^\circ$ can be
  achieved in $\sim$4~seconds with the SCoTSS imager and $\sim$2~minutes with
  the H3D Polaris-H Quad imager as indicated by the vertical and horizontal dashed
  lines.
  Both data curves are assembled from real data acquired from the two
  imagers.  The H3D Polaris-H Quad imager was exposed to a 5~$\times$ higher
  actual field and the times were scaled up to the standard of comparison.  
  The SCoTSS imager was exposed to a 0.5~$\times$ lower actual field and the
  times were scaled down to the standard of comparison.}
\label{fig:TTI}
\end{figure}

In Figure~\ref{fig:TTI} the time-to-image metrics of the SCoTSS single-module 
and $3\times3$-module imagers and of the 
H3D Polaris-H Quad imager when exposed to a gamma-ray flux
equivalent to that emitted by a 1~mCi Cs-137 point source located at a
distance of 10~m are shown. 
All three imagers exhibit a precision of localization which improves rapidly with
increasing acquisition time.
The $3\times3$-module SCoTSS imager achieves a localization precision of better than 
$2^\circ$ in under 3~seconds.
This points to the possibility to make use of the $3\times3$-module SCoTSS
imager not only for isotope alarming, identification and direction-finding
in motion, but also
for imaging in motion.
In contrast to the $3\times3$-module imager, the single-module hand-held
SCoTSS imager would require longer, approximately one minute to achieve
the same image precision.
Despite its superior energy resolution, the H3D Polaris-H Quad imager requires
even longer, over $\sim$2~minutes, to achieve this image precision.
Since operators who require a handheld imager probably do not also wish to
carry a tripod it is worth pointing out that sufficiently rapid imaging using
a handheld imager is only going to be possible for much higher
activity sources, or unshielded sources.

The importance of the comparison in Figure~\ref{fig:TTI} is in the
demonstration of a quantitative comparison of the performance of detectors of
radically different design.
Although some of the designs under comparison have superior efficiency and
others have superior angular
resolution measure and energy resolution, it is nevertheless possible to
compare their overall performance quantitatively, at least for the case of
reconstruction of a point source.
Future presentations of imager performance should include this measure, to allow
an operator to choose between 
sensitivity and volume or cost, according to his or her mission space.

\section{Discussion and Conclusion}
Characteristics of a Compton imager based on crystalline scintillator and read
out with silicon photomultipliers have been presented previously by this
group.
What is new in this study is that we report on the imager's performance from
the particular perspective of a community of responder end users.
Mobile surveyors value the capability to map the second by second counts in
energy windows.
For this they require an imager with not only georeferenced spectra but
sufficient sensitivity for detection and alarming based on the information gained in one second of
acquisition while moving.
Operators also value obtaining an indication of the direction to the source in
real time while in motion.
For an imager, this again means very high sensitivity as well as omni-directional acceptance.
The self-shielding information any imager possesses can be used to rapidly produce
an approximate direction to the
hottest source in the full 4$\pi$ solid angle.

It can be difficult for operators to choose among detectors of 
differing technologies when for example one detector provides
superior energy resolution while another detector has
superior efficiency.
Operators may
have to choose between energy resolution and sensitivity as the materials
which provide superior energy resolution tend to be too expensive to acquire
in large quantities.
Some operators nevertheless require the imaging cability to be provided in as
small and light a package as possible.
The time-to-image quantity discussed in this manuscript can provide a means
for quantitative comparison of the overall performance of imagers which have 
different strengths on individual parameters.
Publication of this measure by the various groups with imagers under
development should be performed.
Armed with this quantitative comparison of imager performance, operators
could then balance the competing requirements for reduced size and complexity
against improved efficiency and resolution for their particular mission space.

\section*{Acknowledgements}
The authors are grateful to Dr.~C.~Wahl from H3D for helpful discussions and 
for provision of the data from the Polaris-H Quad detector shown herein.
The authors gratefully acknowledge helpful discussions with a number of 
scientists from Defense Research and Development Canada.
This work has been made possible with funding from Defense Research and
Development Canada's Canadian Safety and Security Program.
This is NRCan Contribution number 20180164.

\end{document}